# Reconstruction of purely absorbing, absorbing and phase-shifting, and strong phase-shifting objects from their single-shot in-line holograms

Tatiana Latychevskaia* and Hans-Werner Fink

Physics Institute, University of Zurich

Winterthurerstrasse 190, 8057 Zurich, Switzerland

*Corresponding author: tatiana@physik.uzh.ch

**ABSTRACT**

We address the problem of reconstructing phase-shifting objects from their single shot in-line holograms. We show that a phase-shifting object cannot be reliably recovered from its in-line hologram by non-iterative reconstruction routines, and that an iterative reconstruction should be applied. We demonstrate examples of simulated in-line holograms of objects with the following properties: purely absorbing, both absorbing and phase shifting, and strong phase-shifting. We investigate the effects of noise and contrast in holograms on the reconstruction results and discuss details of an optimal iterative procedure to quantitatively recover the correct absorbing and phase-shifting properties of the object. We also review previously published reconstructions of experimental holograms and summarize the optimal parameters for retrieval of phase-shifting objects from their in-line holograms.

## 1. INTRODUCTION

For demonstrating the holography principle, Dennis Gabor performed his first experiment by using a simple transparency with the written words "Huygens, Young, Fresnel" forming an opaque object [1-3]. Since then, various objects have been imaged by digital in-line holography, including non-opaque weak phase-shifting objects [4-5]. However, there is a concern that phase-shifting objects cannot be correctly reconstructed from their in-line holograms. Phase-shifting objects are easily imaged by off-axis holography [6] or by other techniques, such as laser probe beam reshaping [7-8]. Koren *et al.* were the first to address the problem by simulations and reconstructions of in-line holograms of a one-dimensional weak phase-shifting object exhibiting a phase shift of up to 0.12 radians [9] and a two-dimensional weak phase-shifting object causing a phase shift of up to 0.07 radians [10]. They were able to show that the result of a reconstruction of a weak phase-shifting object by an iterative routine depends on the Fresnel number of the setup $N_F = \frac{(\text{maximum object size})^2}{\lambda z}$, where $\lambda$ is the wavelength and $z$ is the diffraction distance; i.e. object to screen separation. The Fresnel number must be sufficiently low; for example, at $N_F = 10$ an object was successfully reconstructed but at $N_F = 100$ its reconstruction was not reliable. In this work we show that phase-shifting objects can be reconstructed from their in-line holograms provided that the Fresnel number and the ratio of object size to total illuminated area is sufficiently low. We discuss why phase-shifting objects are difficult to reconstruct, study the effect of noise and contrast in holograms, and summarize the conditions under which phase-shifting objects can quantitatively and correctly be reconstructed. Finally, we compare our findings with reported reconstructions of experimental holograms.

## 2. SIMULATED HOLOGRAMS

In general, any object can be described by its transmission function $t(x, y)$ [11-13]:

$$t(x, y) = e^{-a(x,y)} e^{i\varphi(x,y)} = t_0(x, y) \cdot e^{i\varphi(x,y)} \tag{1}$$

where $t_0(x, y)$ is a real-valued function which corresponds to the transmittance of the object, and $a(x, y)$ and $\varphi(x, y)$ describe the absorbing and phase-shifting properties of the object, respectively. Where no object is present, $t(x, y) = 1$ and the passing wave forms the reference wave. Here, $(x, y)$ are the coordinates in the object plane.

The holograms are simulated in an in-line holography scheme using plane wave, as shown in Fig. 1.

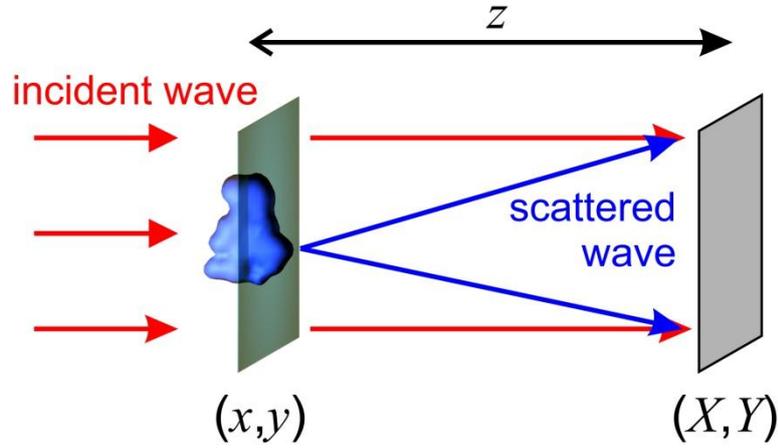

Fig. 1. In-line holography scheme realized with plane waves.

For the case of an in-line hologram recorded with spherical waves, the hologram can be reconstructed as if it were recorded with plane waves under geometrical size recalculations as described in [13]. We therefore limit our study to holograms obtained with plane waves, as a similar analysis is applicable to in-line holograms recorded with spherical waves. The object area is set to 2 × 2 cm$^2$ and the wavelength to 532 nm. To select the object size we applied the following consideration. According to Gabor, not more than about 1% of the illuminated field should be covered by the sample [3]. We introduce the parameter σ = (area occupied by object)/(total illuminated area). As an object we selected a sphere of 1 mm in diameter, which corresponds to σ = 2·10$^{-4}$. The object-to-detector distance is selected as 1 m, thus providing a relatively low Fresnel number, $N_F$ = 1.88, which should result in reliable reconstructions [10].

The angular spectrum theory [14-15] is applied here for the simulation and reconstruction of holograms [13]. Holograms of the sample with the transmission function $t(x, y)$ are simulated by the propagation of the optical field from the object plane towards to the hologram plane:

$$H(X,Y) = \left| \mathrm{FT}^{-1}\left[ \mathrm{FT}\{t(x,y)\} \exp\left( \frac{2\pi i z}{\lambda} \sqrt{1-(\lambda f_x)^2 - (\lambda f_y)^2} \right) \right] \right|^2 \quad (2)$$

where $\mathrm{FT}$ and $\mathrm{FT}^{-1}$ are the Fourier transform and the inverse Fourier transform, respectively, and ($f_x$, $f_y$) are the spatial frequencies.

We simulate and reconstruct in-line holograms of three types of objects:

Type A: a purely absorbing object with a minimum transmittance of 0.8, shown in Fig. 1;

Type B: an object with absorption (minimum transmittance of 0.6) and phase-shifting properties (phase shift of up to 2 radians), shown in Fig. 2;

Type C: a strong phase-shifting object with a minimum transmittance of 0.9 and phase shift of up to 4 radians, shown in Fig. 3.

For all three types of objects, it is assumed that they are not entirely transparent; that is, the transmittance of the object is assumed to be less than 1, as any realistic object possesses some absorption.

The complex-valued transmission function of the sample $t(x, y)$ is reconstructed by the propagation of the optical field from the detector plane backwards to the object plane:

$$t(x,y) = \mathrm{FT}^{-1}\left[ \mathrm{FT}\{H(X,Y)\} \exp\left(-\frac{2\pi i z}{\lambda}\sqrt{1-(\lambda f_x)^2-(\lambda f_y)^2}\right)\right]. \qquad (3)$$

The mismatch between original transmittance $t_0(m,n)$ or phase-shifting distribution $\varphi_0(m,n)$ and the recovered distributions $t(m,n)$ and $\varphi(m,n)$, respectively, is estimated by calculating the error:

$$E_t = \frac{1}{\mathrm{N}^2} \sum_{m,\,n\,=\,1...\mathrm{N}} |t_0(m,n) - t(m,n)|, \quad E_\varphi = \frac{1}{\mathrm{N}^2} \sum_{m,\,n\,=\,1...\mathrm{N}} ||\varphi_0(m,n)| - |\varphi(m,n)||, \qquad (4)$$

where $m, n = 1...N$ are the pixel numbers and the summation is performed over all pixels. For example, the mismatch between object type A and type C estimated in this way results in $E_t = 9.18 \cdot 10^{-4}$ and $E_\varphi = 5.24 \cdot 10^{-3}$. These values can be used as thresholds indicating that two compared distributions are completely different.

## 3. NON-ITERATIVE RECONSTRUCTION

The simulated holograms of the three types of objects and their non-iterative reconstructions are shown in Figs. 2–4.

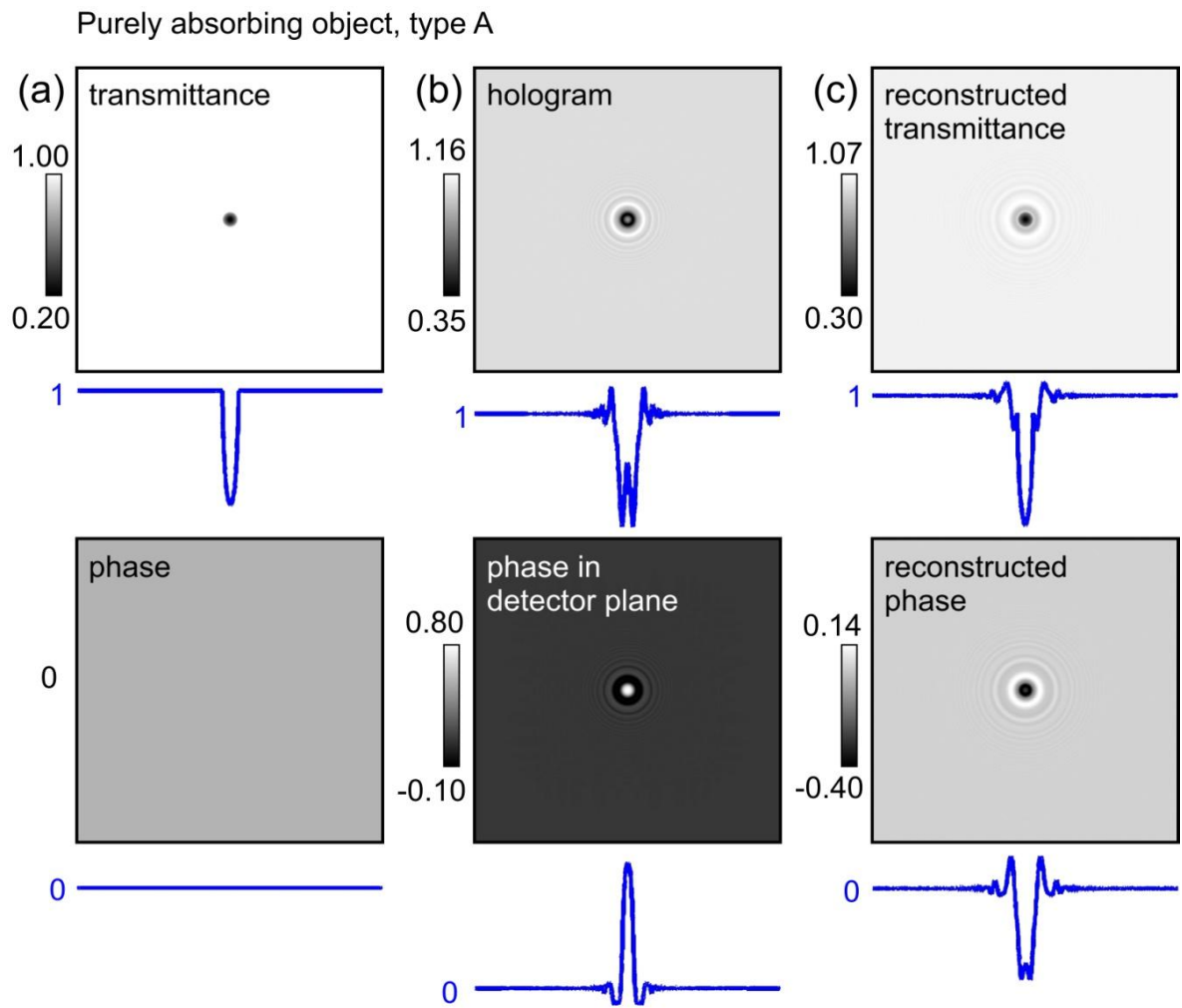

Fig. 2. **Purely absorbing object, type A**. (a) Distributions of transmittance (top) and phase (bottom) of the object. (b) Simulated hologram (top) and the phase distribution at the detector plane (bottom). (c) Reconstructed transmittance (top) and phase (bottom) distributions. The blue curves are the line scans through the centers of the corresponding images.

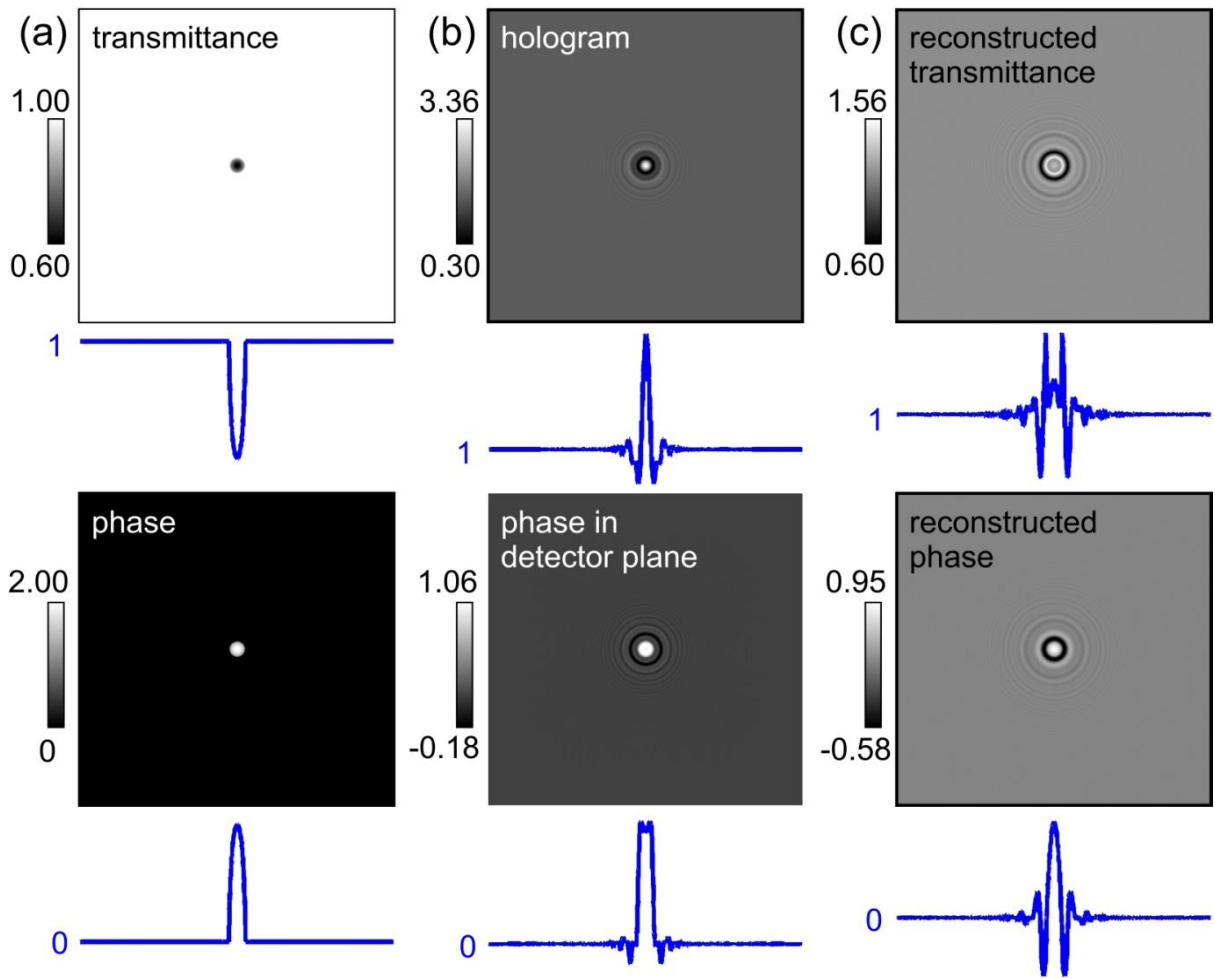

Fig. 3. **Object with absorbing and phase-shifting properties, type B**. (a) Distributions of transmittance (top) and phase (bottom) of the object. (b) Simulated hologram (top) and phase distribution at the detector plane (bottom). (c) Reconstructed transmittance (top) and phase (bottom) distributions. The blue curves are the line scans through the centers of the corresponding images.

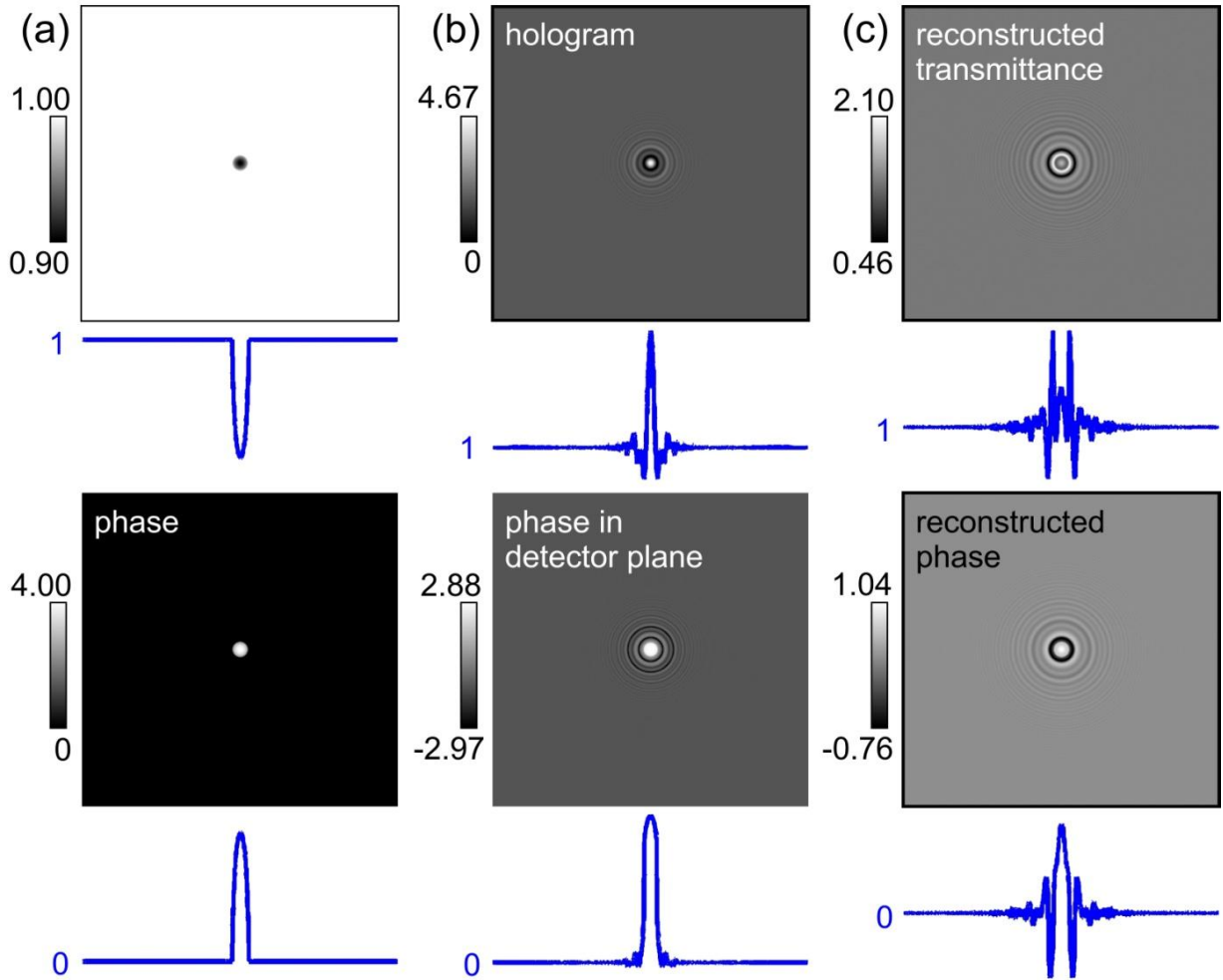

Fig. 4. **Object with strong phase-shifting properties, type C**. (a) Distributions of transmittance (top) and phase (bottom) of the object. (b) Simulated hologram (top) and the phase distribution at the detector plane (bottom). (c) Reconstructed transmittance (top) and phase (bottom) distributions. The blue curves are the line scans through the centers of the corresponding images.

The following conclusions can be drawn about the results shown in Figs. 2–4.

**Purely absorbing object, type A:** The reconstructed transmittance distribution qualitatively resembles the initial transmittance distribution. However, the reconstructed phase distribution is non-zero and might lead to a wrong conclusion that the object possesses some phase-shifting properties. The estimated errors are $E_t = 3.32 \cdot 10^{-3}$ and $E_\varphi = 3.81 \cdot 10^{-3}$.

**Object with absorbing and phase-shifting properties, type B:** Both reconstructed distributions exhibit artificial rings, which makes it difficult to be confident that the object is reconstructed at the correct in-focus position. This can be a problem when reconstructing an experimental hologram, where the exact in-focus position is not known. Moreover, neither absorption nor phase distributions are reconstructed correctly. This implies that a realistic object which possesses absorption and phase-shifting parts cannot be correctly reconstructed from its in-line hologram. The estimated errors are $E_t = 7.51 \cdot 10^{-3}$ and $E_\varphi = 9.15 \cdot 10^{-3}$.

**Object with strong phase-shifting properties, type C:** (1) Holograms of strong phase objects exhibit higher contrast fringes compared to holograms of purely absorbing objects. Some of the fringes are extremely bright, and the maximum intensity in the normalized hologram reaches 4.67 in arbitrary units. (2) The phase distribution at the detector plane resembles the strong phase introduced by the object in the object plane. This means that once the wavefront of the passing wave has been changed by a strong phase-shifting object, the phase of the wavefront has acquired the slow-varying large component, which is preserved during the wavefront propagation to the detector. When the hologram is recorded, this slow-varying phase component, and with it the information about the object phase-shifting properties, is lost. This is exactly the reason why it is difficult to recover the phase distribution of a strong phase-shifting object from its in-line hologram. (3) As in the case of a type B object, it is difficult to judge if the object is reconstructed at the correct in-focus position since the reconstruction reveals a superposition of concentric rings attributed to both the reconstructed object and its twin image. The estimated errors are large: $E_t = 1.36 \cdot 10^{-2}$ and $E_\varphi = 1.79 \cdot 10^{-2}$. Attempts to simulate and reconstruct an object with an even higher phase shift produce similar results. However, it is worth noting that when the maximal phase shift was set to 7 radians, the phase distribution was perfectly reconstructed. This anomaly however only confirms that for general phase-shifting objects of unknown phase shift, a non-iterative reconstruction cannot deliver reliable results.

In 2012, Jericho et al. published a set of phase-shifting objects reconstructed from their experimental in-line holograms recorded with spherical waves [16]. The objects were immersed into a media with a comparable refractive index to minimize the phase shift, and the thickness of the objects was estimated by evaluating this small phase shift. Although all the reconstructions were obtained by a non-iterative procedure, for some of the objects good agreement between the evaluated and the expected object thickness was achieved.

In summary, only a rough estimation of the object distribution can be obtained by performing a non-iterative reconstruction. In the case of a strong phase-shifting object, it is even difficult to assign the correct in-focus position of the reconstructed object. For all three types of objects, the estimated errors are higher than their threshold values, demonstrating large mismatches between the reconstructed and original distributions.

## 4. ITERATIVE RECONSTRUCTION

Figure 4(b) shows that the phase distribution in the detector plane is strongly dominated by the phase shift introduced by the object. As a detector records only an intensity distribution, the phase and therefore the signature of the object phase-shifting properties are lost. This resembles the phase problem in coherent diffractive imaging [17-18], which is conventionally solved by iterative phase retrieval methods [19]. Such iterative approach has already been demonstrated for the reconstruction of phase-shifting objects from their in-line holograms [10-11, 20-24]. When performing an iterative reconstruction, the requirement for the ratio of $\sigma$ can be relaxed from the value of $\sigma<0.01$ proposed by Gabor [3] to $\sigma<0.25$, as required by the oversampling condition in coherent diffractive imaging [17]. In this section we describe the details of the algorithm that allows for the reconstruction of the three types of objects from their in-line holograms and we summarize the results of the reconstruction of objects with different absorbing and phase-shifting properties.

The iterative reconstruction scheme is shown in Fig. 5.

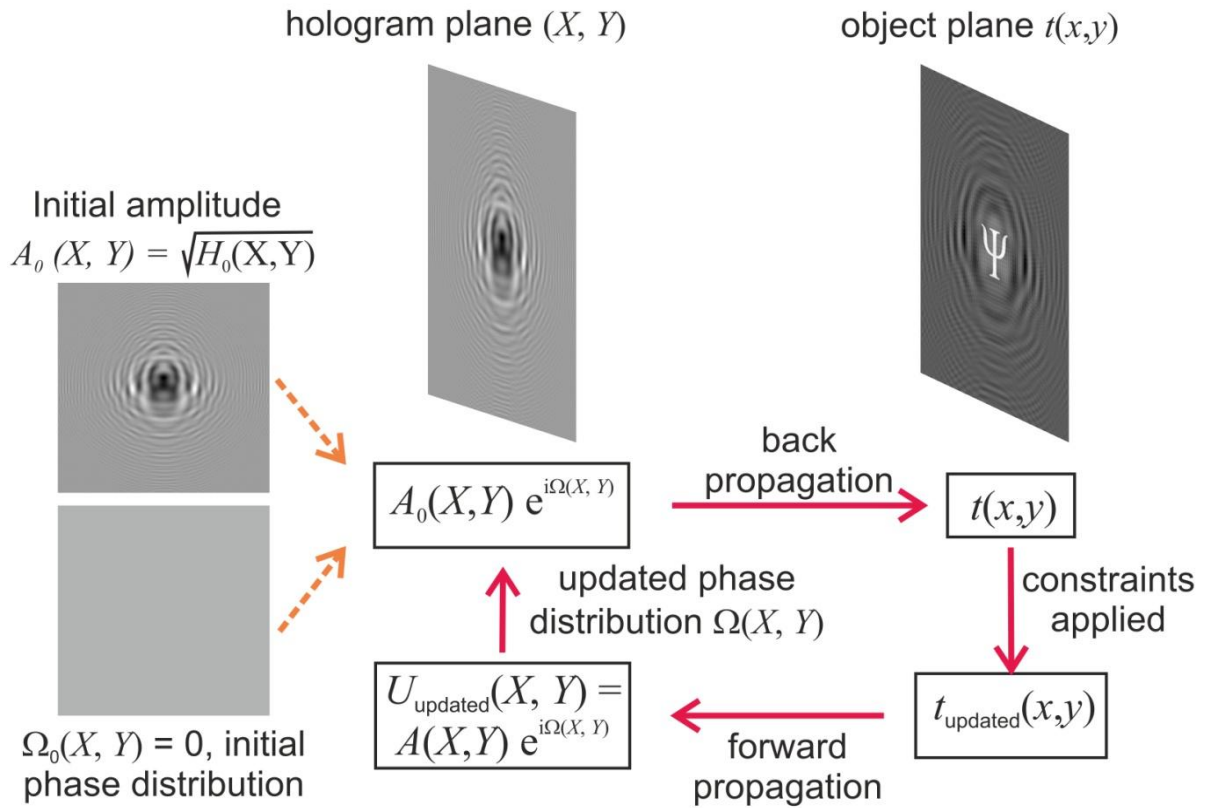

Fig. 5. Iterative reconstruction scheme. $H_0(X, Y)$ is the normalized hologram, that is the hologram divided with the background image.

The iterative procedure begins with the formation of the wavefront distribution in the detector domain. The amplitude of the initial wavefront at the detector plane is obtained by taking the square root from the hologram intensity, and the initial phase distribution is set to zero. We have observed that a reconstruction with initial random phase instead of zero phase eventually leads to the same result.

(i) The wavefront at the detector plane $U_H(X,Y)$ is propagated to the object domain, where the transmission function $t(x,y)$ is extracted:

$$t(x,y) = \mathrm{FT}^{-1}\left[\mathrm{FT}\{U_H(X,Y)\}\exp\left(-\frac{2\pi i z}{\lambda}\sqrt{1-(\lambda f_x)^2-(\lambda f_y)^2}\right)\right]. \tag{5}$$

(ii) The transmission function of the object is subject to the following constraints. For all iterations, it is multiplied by a support function [10, 20, 22], which sets the values of the transmission function to 1 outside the support and leaves them unchanged inside the support. The support distribution is obtained from the initial non-iterative reconstruction; it is larger by 7 pixels than the diameter of the object (25 pixels), and the edges of the mask are not sharp but Gaussian blurred (blur = 2 pixels). The constraint of non-negative absorption is applied [11], implying that negative values of absorption are replaced by zeros. This is equivalent to the condition that the transmittance must not exceed 1, and values that exceed 1 are replaced by 1. The object phase-shifting distribution is kept unchanged. For all iterations, the updated transmittance distribution is smoothed by convolution with a filter of a 5 × 5 pixels kernel in the form of a Gaussian distribution

$$K = \begin{pmatrix} 1 & 1 & 1 & 1 & 1 \\ 1 & 1 & 1 & 1 & 1 \\ 1 & 1 & 4 & 1 & 1 \\ 1 & 1 & 1 & 1 & 1 \\ 1 & 1 & 1 & 1 & 1 \end{pmatrix} \tag{6}$$

in order to smooth the interference pattern inside the object distribution and to suppress the accumulation of noisy peaks. The transmission function is then updated, leading to $t_{updated}(x,y)$.

(iii) The wavefront is propagated from the object to the detector plane by computing:

$$U_{updated}(X,Y) = \left|\mathrm{FT}^{-1}\left[\mathrm{FT}\{t_{updated}(x,y)\}\exp\left(\frac{2\pi i z}{\lambda}\sqrt{1-(\lambda f_x)^2-(\lambda f_y)^2}\right)\right]\right|^2. \tag{7}$$

(iv) The phase distribution of the updated wavefront is adapted for the next iteration starting at (i), while the amplitude distribution is always replaced by the square root of the measured intensity.

The reconstruction quickly converges and stagnates after a few tens of iterations out of a total of 300 iterations performed. The results of the iterative reconstructions are shown in Figs. 6 and 7.

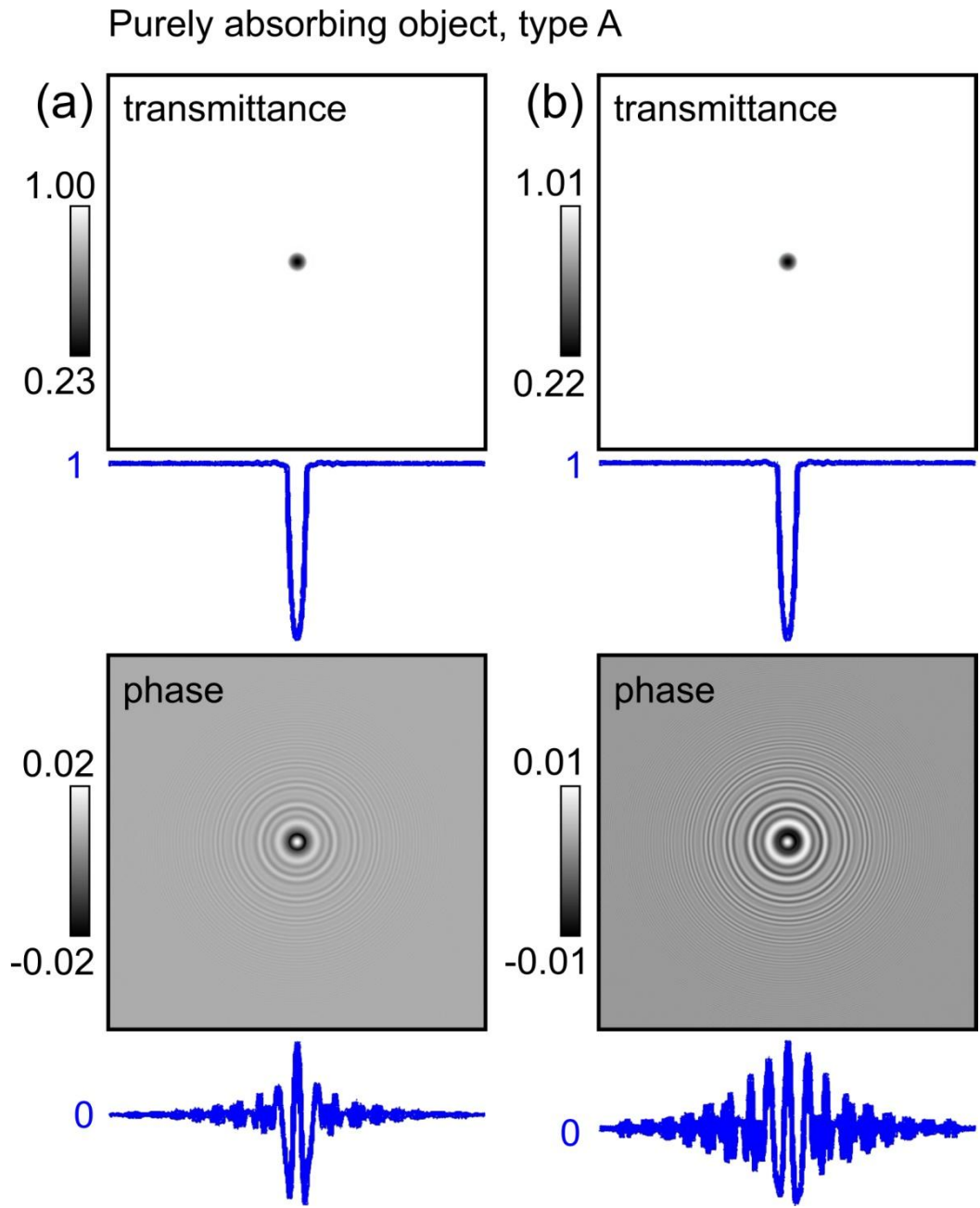

Fig. 6. Iteratively reconstructed transmittance (top row) and phase (bottom row) distributions of **purely absorbing object, type A**. (a) No constraint is superimposed onto the phase distribution in the object domain during the iterative procedure. (b) The same as (a) but with the constraint that the phase distribution in the object domain is set to zero. The blue curves are line scans through the centers of the corresponding images.

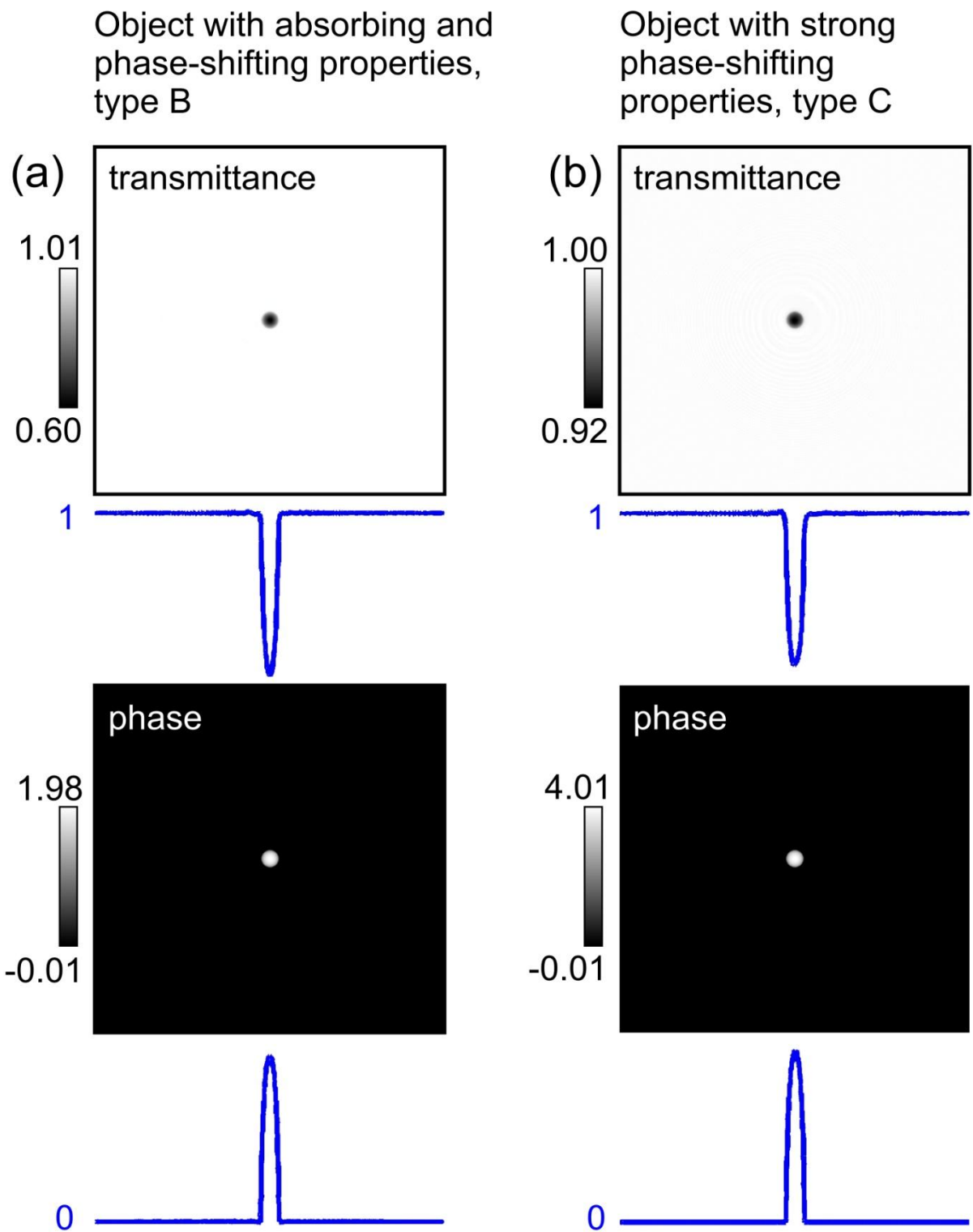

Fig. 7. Iteratively reconstructed transmittance (top) and phase (bottom) distributions of the object with **(a) absorbing and phase-shifting properties, type B** and **(b) strong phase-shifting properties, type C**. The blue curves are the line scan through the center of the corresponding image.

The following conclusions can be drawn by visual inspection of the reconstructions shown in Figs. 6–7.

**Purely absorbing object, type A,** shown in Fig. 6: As in the case of non-iterative reconstructions, the reconstructed transmittance distribution matches the original transmittance distribution well and almost reaches the predefined minimum of 0.2. The reconstructed phase-shifting distribution is very close to but different from zero. The calculated errors are smaller than in the case of non-iterative reconstructions: $E_t = 8.08 \cdot 10^{-4}$ and $E_\varphi = 7.58 \cdot 10^{-4}$. When the additional constraint of setting the phase-shifting distribution to zero is applied in the object plane, the recovered object phase gets smaller and has values within (–0.01, 0.01); see Fig. 6(b). The calculated errors are $E_t = 7.94 \cdot 10^{-4}$ and $E_\varphi = 7.22 \cdot 10^{-4}$.

**Object with absorbing and phase-shifting properties, type B**, shown in Fig. 7(a), and object with **strong phase-shifting properties, type C**, shown in Fig. 7(b)**:** For both types of objects, the transmittance and phase-shifting distributions are almost perfectly recovered. The calculated errors are

Type B: $E_t = 5.63 \cdot 10^{-4}$ and $E_\varphi = 5.89 \cdot 10^{-4}$;

Type C: $E_t = 2.44 \cdot 10^{-4}$ and $E_\varphi = 2.41 \cdot 10^{-4}$.

By comparing the reconstructions obtained by non-iterative with those obtained by iterative procedures, it can be concluded that when an object possesses strong phase-shifting properties, its phase distribution cannot be correctly recovered by conventional, non-iterative reconstruction of the hologram, and an iterative routine must be applied. The reconstruction errors for all three types of objects are relatively small, indicating that the retrieved distributions are close to the original ones.

## 5. EFFECT OF SIGNAL-TO-NOISE RATIO IN HOLOGRAMS ON RECONSTRUCTION

Until now we have considered ideal simulated holograms that are free from noise. In this section we address more realistic holograms and study two factors: (1) noise in holograms, quantitatively measured by the signal-to-noise ratio (SNR), and (2) intrinsic resolution of holograms, given by the resolution of the optical system, the numerical aperture, mechanical vibration and possible other resolution limiting factors. We study the effects of these two factors on a hologram of a type C object.

To study the effect of noise on the hologram reconstruction, Gaussian distributed noise of pre-defined SNR = 5, 10, and 20 is superimposed onto an ideal hologram. The results of the iterative reconstructions are shown in Fig. 7. It can be concluded that noise severely affects the quality of the reconstructed transmittance distribution, while the phase distribution is almost perfectly quantitatively reconstructed for a SNR = 5 – 20. The calculated errors are extremely large:

SNR = 5: $E_t = 5.74 \cdot 10^{-2}$ and $E_\varphi = 5.80 \cdot 10^{-2}$

SNR = 10: $E_t = 2.82 \cdot 10^{-2}$ and $E_\varphi = 2.83 \cdot 10^{-2}$

SNR = 20: $E_t = 1.41 \cdot 10^{-2}$ and $E_\varphi = 1.41 \cdot 10^{-2}$.

A hologram with reduced contrast was created by blurring the ideal hologram by convolution with a filter of a 5 × 5 pixels kernel in the form of a Gaussian distribution, given by Eq. 6. In addition, Gaussian distributed noise was superimposed onto the hologram leading to SNRs = 5, 10, and 20 as in the example above; the results are shown in Fig. 9. The iterative routine here was optimized in the following manner: the constraint of positive absorption was applied only for the first 100 iterations out of 300 iterations, and for the remaining iterations, the updated transmittance distribution was smoothed by convolution with the kernel given by Eq.7 for 5 times after each iteration. Reconstructions obtained from the blurred holograms are qualitatively better than those obtained from non-blurred holograms; compare Figs. 8 and 9. The calculated errors are again large:

SNR = 5: $E_t = 5.36 \cdot 10^{-2}$ and $E_\varphi = 5.49 \cdot 10^{-2}$

SNR = 10: $E_t = 2.68 \cdot 10^{-2}$ and $E_\varphi = 2.77 \cdot 10^{-2}$

SNR = 20: $E_t = 1.39 \cdot 10^{-2}$ and $E_\varphi = 1.46 \cdot 10^{-2}$.

In practice, noise in holograms can be reduced by averaging a sequence of holograms acquired over a certain time period, while the resolution of a hologram can be enhanced in some cases by recording and subpixel alignment of shifted holograms [25].

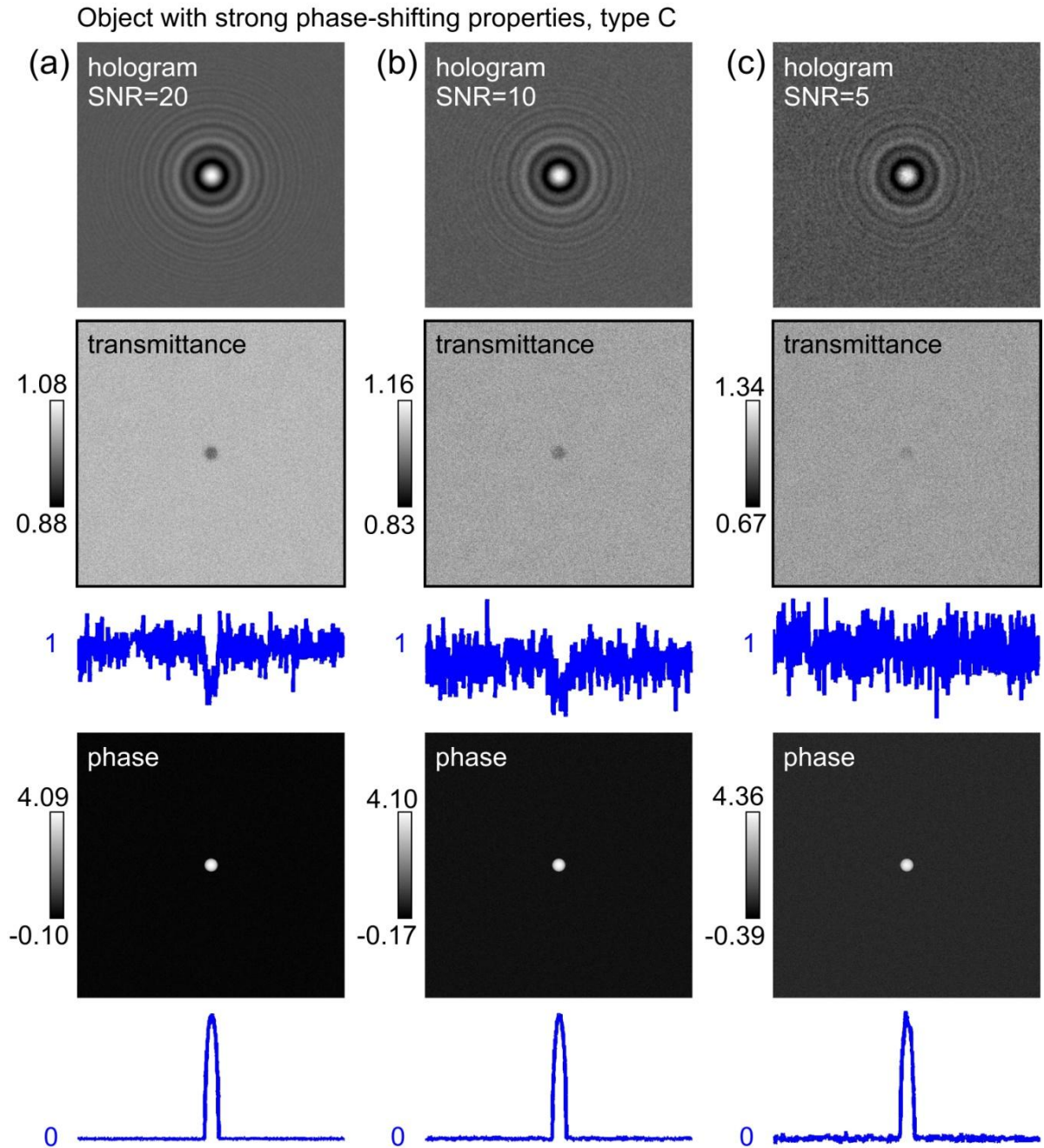

Fig. 8. **The effect of noise added to a hologram of an object with strong phase-shifting properties, type C**. Distributions of the central 500 × 500 pixels region of the hologram (top row); reconstructed transmittance (middle row) and phase (bottom row) distributions. The results for (a) SNR = 20, (b) SNR = 10, and (c) SNR = 5 are shown. The blue curves are the line scans through the centers of the corresponding images.

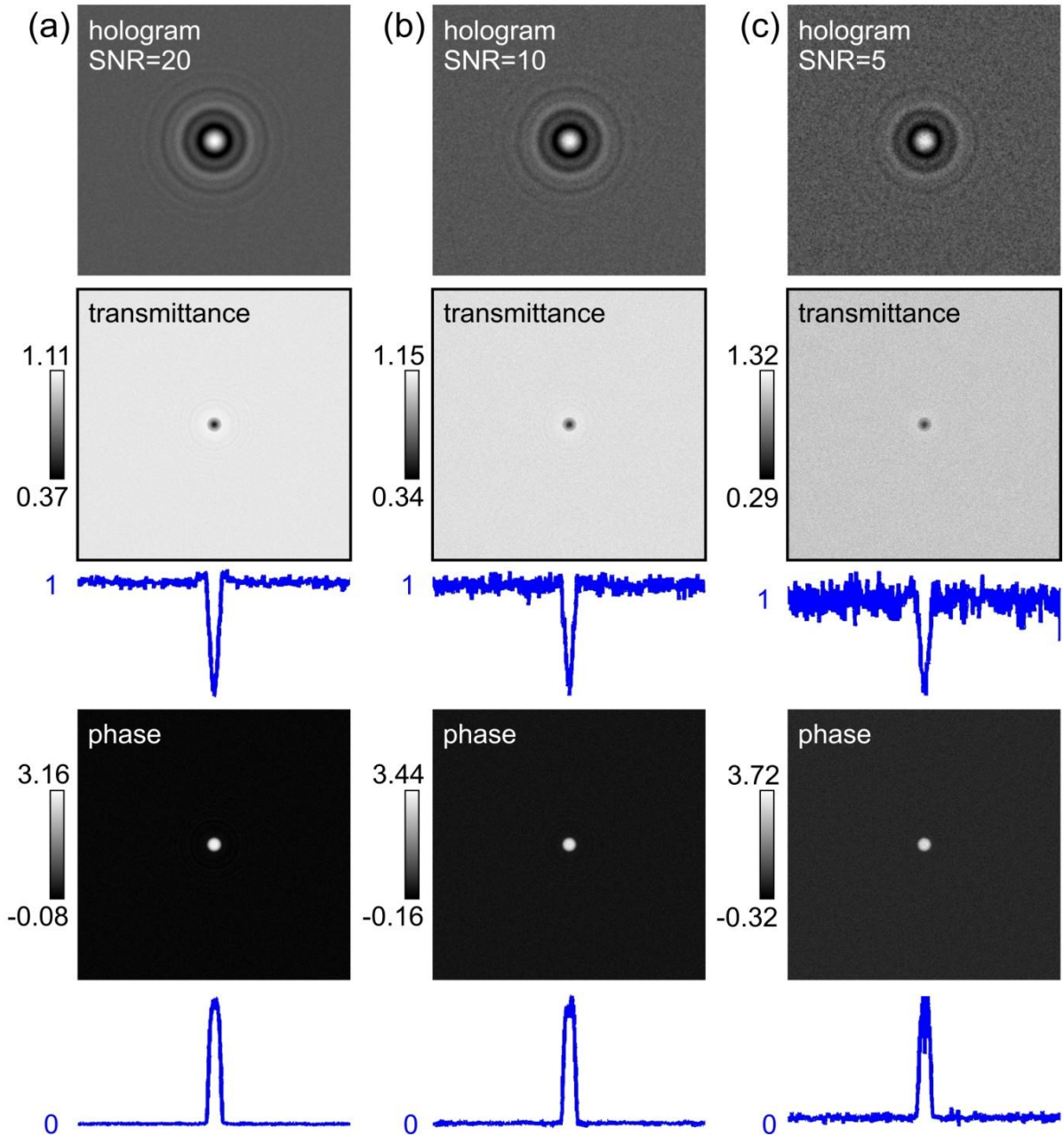

Fig. 9. **The effect of blurring of a hologram of an object with strong phase-shifting properties, type C**. Distributions of the central 500 × 500 pixels region of the hologram (top row); reconstructed transmittance (middle row) and phase (bottom row) distributions. The results for (a) SNR = 20, (b) SNR = 10, and (c) SNR = 5 are shown. The blue curves are the line scans through the centers of the corresponding images.

## 6. EFFECT OF THE OBJECT SIZE

In this section we reconstruct holograms simulated at the same parameters as above, but for a larger object, a sphere of diameter of 7 mm, so that the ratio σ = 0.1 is ten times larger than required for a successful hologram reconstruction according to the Gabor criterion. We simulated holograms for the three aforementioned types of objects. Their non-iterative reconstructions are shown in Fig. 10 indicating that for no type of object a correct reconstruction was achieved. Only for purely absorbing object of type A, the reconstructed absorption distribution approaches the correct distribution. Quantitatively, the calculated errors are large:

Type A: $E_t = 2.42 \cdot 10^{-2}$ and $E_\varphi = 1.44 \cdot 10^{-2}$

Type B: $E_t = 2.37 \cdot 10^{-2}$ and $E_\varphi = 0.138$

Type C: $E_t = 3.43 \cdot 10^{-2}$ and $E_\varphi = 0.278$

Qualitatively, the obtained reconstructions are worse than the reconstructions of a 1 mm sized object, compare Fig. 10 and Figs. 2 – 4. Thus, the parameter σ, or the object size compared to the total illuminated area must be sufficiently small, as already predicted by Gabor.

The iteratively obtained reconstructions are shown in Fig. 11, and they are clearly superior to the non-iterative reconstructions shown in Fig. 10. In particular, for object B and C, the transmittance and phase distributions are better matching the original distributions. The iterative routine for reconstructing object A is as described above: 300 iterations where the positive absorption constraint is applied and the updated transmittance distribution is smoothed by convolution with the kernel given by Eq. 7 for 5 times after each iteration; an additional constraint of setting the phase-shifting distribution to zero is applied in the object plane for all iterations. The iterative routine for reconstructing objects B and C is run for 20000 iterations where the positive absorption constraint is applied for the first 10000 iterations, and for the remaining iterations, the updated transmittance distribution was smoothed by convolution with the kernel given by Eq. 7 for 5 times after each iteration. The support in the objet domain was 4 pixels larger than the contour of the object, and it was smoothed by applying a convolution with the kernel given by Eq. 7.

Quantitatively, the calculated errors are relatively small:

Type A: $E_t = 1.25 \cdot 10^{-3}$ and $E_\varphi = 1.13 \cdot 10^{-3}$

Type B: $E_t = 1.93 \cdot 10^{-3}$ and $E_\varphi = 3.57 \cdot 10^{-2}$

Type C: $E_t = 5.35 \cdot 10^{-4}$ and $E_\varphi = 6.32 \cdot 10^{-3}$.

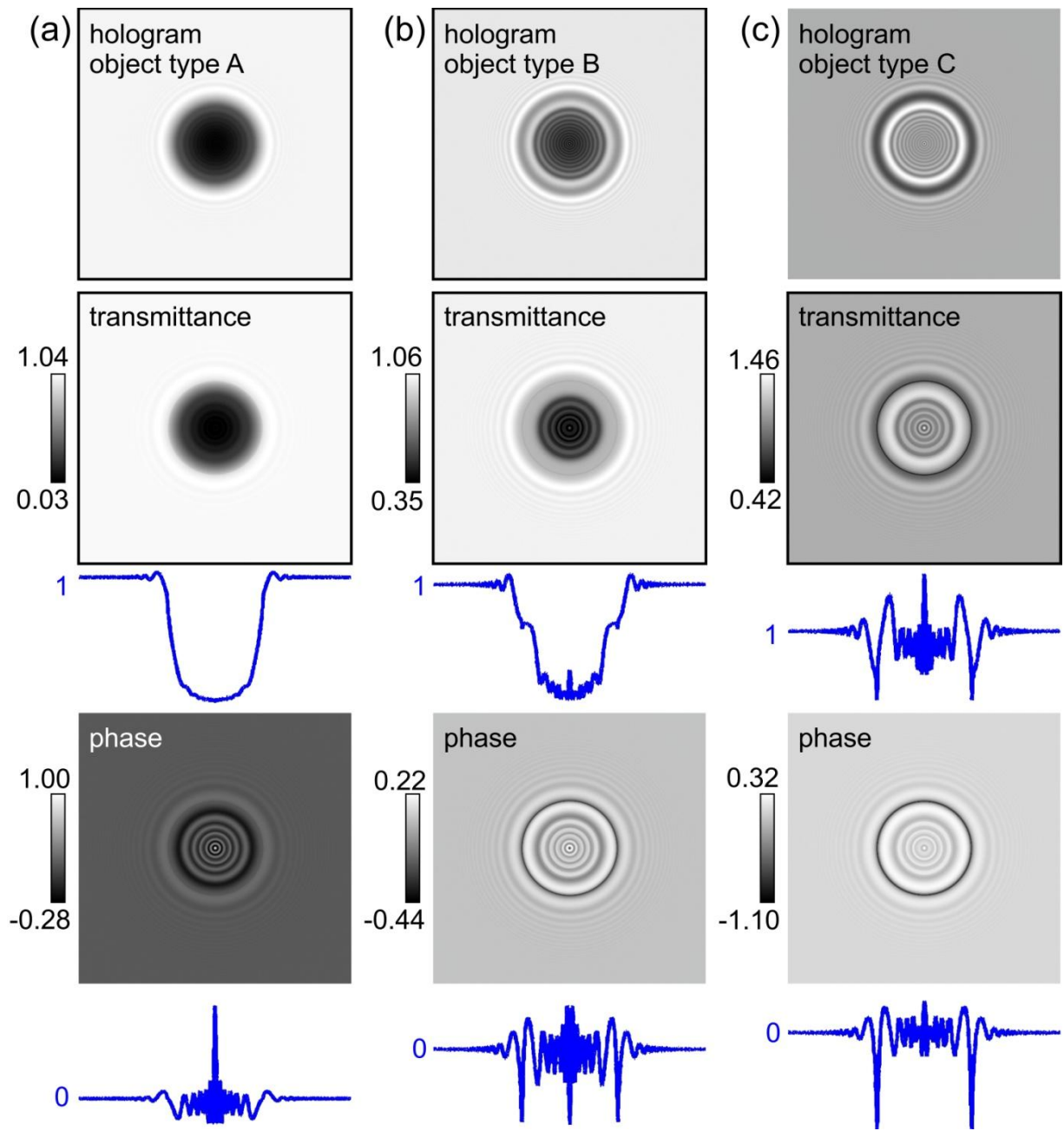

Fig. 10. **The effect of the object size.** Holograms and their reconstructions for large objects of (a) type A, (b) type B, and (c) type C.

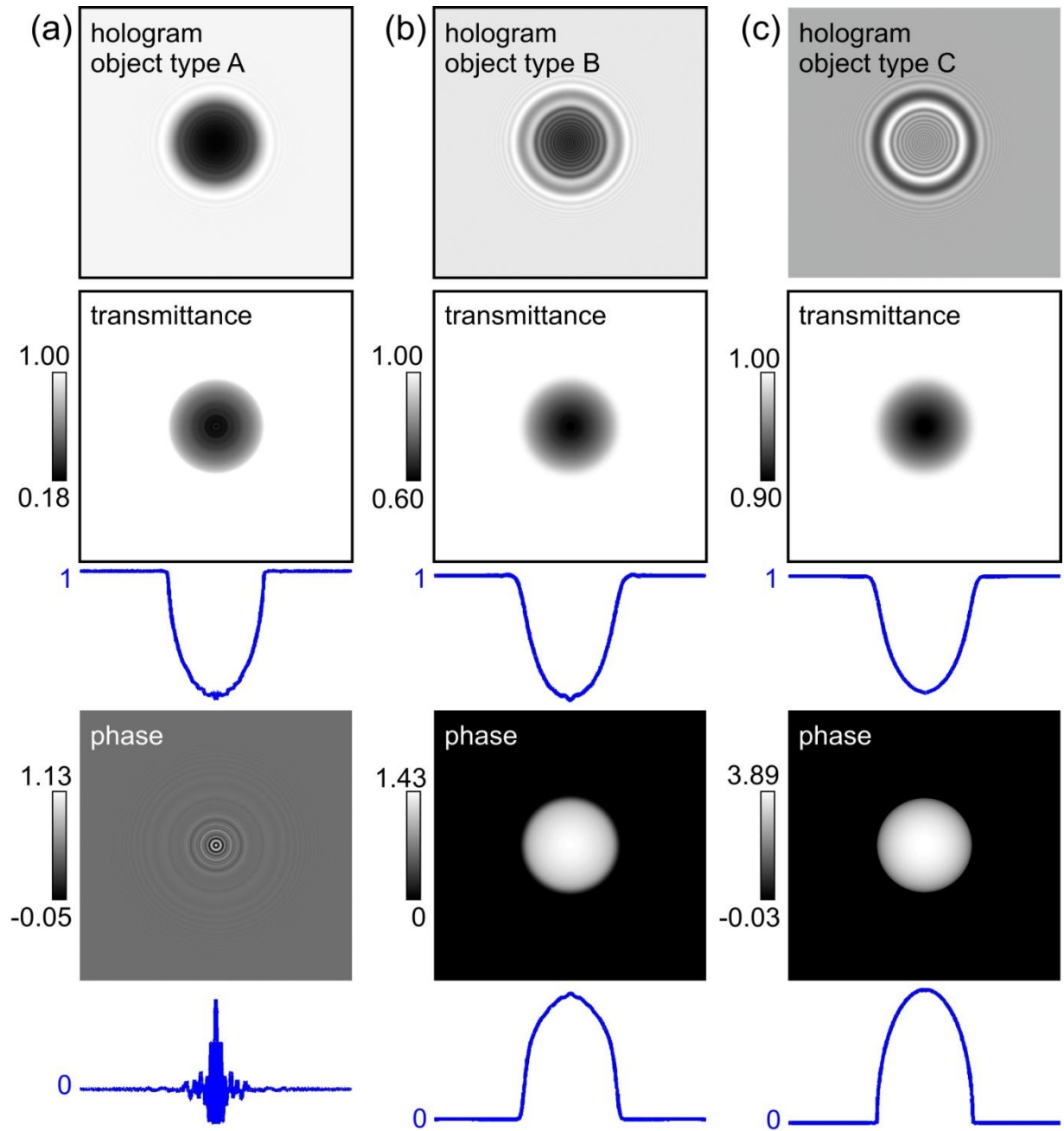

Fig. 11. **The effect of the object size.** Holograms and their iterative reconstructions for large objects of (a) type A, (b) type B, and (c) type C.

## 7. COMPARISON WITH THE RECONSTRUCTION OF EXPERIMENTAL HOLOGRAMS

In a number of publications, simultaneous reconstruction of absorption and phase-shifting properties of objects from their experimental single-shot in-line holograms has been reported [10-11, 16, 20-24, 26]. However, only a few of the published articles provide the possibility to quantitatively cross-validate the iteratively reconstructed phase-shifting distributions with their true values.

Individual latex spheres, 1 μm in diameter, were reconstructed from their in-line holograms by the iterative procedure described above [11], with a recovered maximal phase shift of about 0.35 radians. The expected phase shift however amounts to $\Delta\varphi = \frac{2\pi}{\lambda} D \cdot (n - n_0) \approx 7$ radians, given the refractive index of latex being $n$ = 1.59, thus making such spheres in air a strong phase-shifting object. The hologram parameters of $N_F$ = 0.007 and $\sigma$ = 9.7·10$^{-5}$ were sufficient to obtain reliable reconstructions. In accordance with our simulation study presented above, the mismatch between the values of the theoretical and reconstructed phase-shift can be explained by the very high level of noise of about a SNR = 13 and the low contrast of the interference fringes of the holograms of individual spheres.

Another example of a reconstruction of a strong phase-shifting object from its experimental in-line hologram dealt with a latex nanosphere, 204 nm in diameter, iteratively reconstructed from its high-energy electron in-line hologram [20]. The recovered phase-shift distribution was cross-validated by the phase-shifting distribution reconstructed from the off-axis hologram of the same object. From both holograms, the maximum of the recovered phase-shift of about 14 radians was confirmed. The hologram parameters provide $N_F$ = 72.3 and $\sigma$ = 0.04, were $N_F$ is higher than required to obtain reliable reconstructions. This leads to the conclusion that $N_F$ is a less crucial parameter than $\sigma$. The successful reconstruction of the phase-shifting properties can be explained by (a) very low noise in the hologram, with a SNR = 80, (b) extra smoothing applied to the updated object absorption distribution during the iterative reconstruction, and (c) the large number of iterations: 4000.

In 2012, Schwenke et al. reported on a weak phase-shifting object iteratively reconstructed from its in-line hologram, whereas the maximum of the reconstructed phase of 0.57 radians corresponded to the maximum of the expected phase shift of 0.51 radians [21]. The Fresnel number of the hologram, $N_F$ = 0.55, was just sufficient to obtain reliable reconstructions.

To summarize, the reported successful reconstructions of phase-shifting objects were achieved from their single-shot holograms with low $N_F$ and $\sigma$ and a high SNR.

## 8. CONCLUSIONS

We have studied the reconstruction of purely absorbing, absorbing and phase-shifting, and strong phase-shifting objects from their single-shot in-line holograms. Our simulations show that a phase-shifting object adds a strong but slow varying component to the wavefront distribution which however is lost when only the intensity is recorded by a detection. Thus, the phase-shifting properties of the object are lost during the hologram detection. This problem is similar to the phase recovery problem in coherent diffraction imaging and can only be solved by applying an iterative routine. We showed that only the iterative reconstruction can deliver qualitatively correct reconstruction of a general type of object from its single-shot in-line hologram. For a phase-shifting object, it is even difficult to find the distance at which the reconstructed object appears to be in focus. To correctly identify the in-focus position, several iterative runs should be performed at different hologram-to-sample distances, and the result with the least error should be selected. For a successful reconstruction, the hologram must have a low Fresnel number $N_F$, more importantly a high σ = (object area)/(total illuminated area) ratio, and also low noise, and a sufficient contrast. The signal-to-noise ratio can be increased by averaging over a sequence of holograms. The resolution can be enhanced by acquiring a sequence of shifted holograms followed by subpixel alignment.